\documentclass{article}
\usepackage[utf8]{inputenc}
\usepackage{amsmath}
\usepackage{amssymb}
\usepackage{slashed}
\usepackage{cite}
\usepackage{graphicx}
\usepackage{color}
\usepackage{xcolor,colortbl}
\usepackage{anysize}
\usepackage{subfiles}
\usepackage[colorlinks=true, urlcolor=blue, linkcolor=blue, citecolor=blue]{hyperref}
\usepackage{cleveref}
\usepackage{makecell}
\usepackage{subfigure}

\marginsize{2.0 cm}{2.0 cm}{2.0 cm}{2.0 cm} 

\definecolor{cosmiclatte}{rgb}{1.0, 0.97, 0.91}
\definecolor{lightcyan}{rgb}{0.88, 1.0, 1.0}

\newcommand*{\olra}{\overleftrightarrow}

\title{Impact of ATLAS
constraints on effective dark matter-standard model interactions with
spin-one mediators}
\author{Fabiola Fortuna$^1$ and Pablo Roig$^1$\\
$^{1}$ \small{Centro de Investigación y de Estudios Avanzados,} \\
\small{Apartado Postal 14-740, 07000, Ciudad de México, México}}

\date{}

\begin{document}

\maketitle

\section*{Abstract}

We complement a previous work \cite{Fortuna:2020wwx} using an EFT framework of dark matter and standard model interactions, with spin-one mediators,  exploring a wider dark matter mass range, up to $6.4$ TeV. We use again bounds from different experiments: relic density, direct detection experiments and indirect detection limits from the search of gamma-ray emissions and positron fluxes. Besides, in this paper we add collider constraints by the ATLAS Collaboration in monojet analysis. Moreover, here we tested our previous results in the light of the aforementioned ATLAS data, which turn out to be the most restrictive forlight dark matter masses (as expected), $m_{\rm DM}<M_Z/2$. We obtain a larger range of solutions for the operators of dimension 5, OP1 and OP4, where masses above $43$ GeV and $30$ GeV (but for the $Z$ resonance region, $\sim (M_Z\pm\Gamma_Z)/2$), respectively, are allowed. In contrast, the operator of dimension 6, OP3, has viable solutions for masses $\gtrsim 190$ GeV. For the combination of OP1\&OP3 we obtain solutions (for masses larger than $140$ or $325$ GeV) that depend on the relative sign between the operators.

\section{Introduction}

Understanding the fundamental nature of Dark Matter (DM) is one of the most compelling
problems in particle physics and cosmology, yet despite years of searching, the identity of DM remains a mystery. A favoured paradigm for the particle nature of dark matter is that of Weakly Interacting Massive Particles (WIMPs) \cite{Dutra,Roszkowski:2017nbc, Schumann:2019eaa}. In this scenario the DM interactions with the Standard Model (SM) are sufficiently weak  to meet the constraints of direct \cite{Fu:2016ega, Aprile:2017iyp, Akerib:2016lao, Behnke:2016lsk, Akerib:2016vxi, Tan:2016zwf, Hanany:2019lle} and indirect detection experiments \cite{Hooper:2010mq, Bulbul:2014sua, Urban:2014yda, Choi:2015ara, Ruchayskiy:2015onc, Ackermann:2015zua, Franse:2016dln, Aharonian:2016gzq, Cui:2016ppb, Aartsen:2016zhm, TheFermi-LAT:2017vmf}, but strong enough to generate the relic abundance inferred from measurements of the cosmic microwave background radiation \cite{Planck:2018vyg}. In absence of any direct DM signal, the generality of the effective field theory (EFT) approach may be advantageous \cite{Belanger:2008sj,Goodman:2010ku,Crivellin:2014gpa,Crivellin:2014qxa,Duch:2014xda,Bhattacharya:2021edh, Barman:2021hhg}, as it only uses the known SM symmetries and degrees of freedom, assuming that the typical energy of all relevant processes lies below the mediator mass. We will follow such an approach by using an effective Lagrangian to parameterize the interactions of the dark sector with the SM, and determine the restrictions imposed by the experimental/observational constraints. The Higgs portal (see ref. \cite{Arcadi:2019lka} and references therein) and neutrino portal cases \cite{Cosme:2005sb, An:2009vq, Falkowski:2009yz, Lindner:2010rr, Farzan:2011ck, Falkowski:2011xh, Heeck:2012bz, Baek:2013qwa, Baldes:2015lka, VaniaIllana, Batell:2017rol, HajiSadeghi:2017zrl, Bandyopadhyay:2018qcv, Berlin:2018ztp, Blennow:2019fhy, Hall:2019rld, Hall:2019ank} have received considerably more attention than the case of spin-one mediators, so we will focus on the latter ---both in the Proca or antisymmetric tensor representations---. 

Here we will continue exploring the phenomenological consequences of the EFT scenario developed in ref. \cite{GonzalezMacias:2015rxl} for the interactions between SM and DM particles with heavy mediators. We have already studied the low energy region, with DM masses under $m_Z/2$ in Ref. \cite{Fortuna:2020wwx}. In that analysis, we found solutions complying with the constraints imposed: Z invisible decay width \cite{Workman:2022ynf}, relic density \cite{Workman:2022ynf}, direct detection limits from Xenon1T \cite{Aprile:2018dbl}, PandaX \cite{Ren:2018gyx}, LUX \cite{Akerib:2018hck}, DarkSide-50 \cite{Agnes:2018oej} and CRESST-III \cite{Abdelhameed:2019hmk}. We also employed indirect detection limits from the search of gamma-ray emissions \cite{Drlica} and positron fluxes \cite{Ibarra:2013zia}. In this work we extend the region of DM masses under analysis, from $50$ GeV up to $6.4$ TeV (slightly less than half the LHC center-of-mass collision energy, as our DM particles need to be pair-produced, accounting for the detection jet energy threshold). We use again the restrictions mentioned above and we add collider constraints by the ATLAS collaboration \cite{ATLAS:2021kxv}. In fact we also tested our previous results from the low energy region against the ATLAS data and further restricted the space of solutions.

The paper is organised as follows: In section \ref{sec:lag} we introduce the EFT that
we are using \cite{GonzalezMacias:2015rxl} and highlight the part interesting for this study and our conventions. Then in section \ref{sec:obslimits} we analyze several observational limits: in subsection \ref{subsec:RD} we check that the observed relic abundance can be reproduced in the different cases, in subsection \ref{subsec:DD} we verify the direct detection bounds are respected; and in subsections \ref{subsec:DSSG-1} and \ref{subsec:AMS-1} we consider the indirect bounds given by dwarf spheroidal satellite galaxies and the positron flux, respectively. After that, in section \ref{sec:CollCons} we include the collider constraints by the ATLAS collaboration; finally the discussion and conclusions are presented in section \ref{sec:Concl-1}.

\section{Effective Lagrangian} \label{sec:lag}

We study interactions between dark matter and standard model particles using an effective field theory approach, where we consider that the heavy mediators that generate the interaction are of spin one. In the dark sector we can have scalars, $\Phi$, fermions, $\Psi$, and vectors, $X$. The mediators are weakly coupled to both sectors, dark and standard, and this information is encoded in the effective coefficients $X_\text{eff}$. We assume that the dark fields transform non-trivially under a symmetry group, $\mathcal{G}_\text{DM}$ (that we do not need to specify), while all SM particles are singlets under this $\mathcal{G}_\text{DM}$, which ensures the stability of the DM particle. Also, all dark fields are singlets under the SM gauge group. The consequence of interactions generated by a mediator are that our operators have the form:

\begin{equation}\label{eq:factorized_op}
	\mathcal{O}=\mathcal{O}_\text{SM}\,\mathcal{O}_\text{dark},
\end{equation}

and we know that $O_\text{dark}$ contains at least two fields because we have assumed that the dark fields transform non-trivially under $\mathcal{G}_\text{DM}$.
In the effective Lagrangian, each term has a factor $1/\Lambda^n$, $n=$dim($\mathcal{O}$)-4 , where $\Lambda$ is of the order of the heavy mediator mass(es). We restricted ourselves to operators of mass  dimension $\leq 6$. In this work, we focus on operators with vector and antisymmetric tensor mediators because the models with scalar and fermion mediators have already been studied extensively \cite{Goodman:2010ku,Duch:2014xda,Racco:2015dxa,Bell:2015sza,DeSimone:2016fbz,Cao:2009uw,Cheung:2012gi,Busoni:2013lha,Buchmueller:2014yoa,Patt:2006fw,VaniaIllana,Lamprea19}.

The Lagrangian we use, is conveniently separated into two parts ($\psi$ stands for SM fermions and $B_{\mu\nu}$ is the $U(1)_Y$ field-strength tensor, universal couplings to the SM fermions are assumed):
\begin{itemize}
\item Terms involving dark fermions ($\Psi$):
\begin{equation} \label{lagpsi}
	\mathcal{L}_\text{eff}^\Psi = \frac{\Upsilon_\text{eff}}{\Lambda} B_{\mu\nu}\bar{\Psi}\sigma^{\mu\nu}\Psi+\frac{A_\text{eff}^{L,R}}{\Lambda^2}\bar{\psi}\gamma_\mu \psi\bar{\Psi}\gamma^\mu P_{L,R}\Psi+\frac{\kappa_\text{eff}^{L,R}}{\Lambda^2}B_{\mu\nu}\bar{\Psi}\left(\gamma^\mu \overleftrightarrow{\mathcal{D}}^\nu - \gamma^\nu\overleftrightarrow{\mathcal{D}}^\mu \right) P_{L,R}\Psi.
\end{equation}
\item Terms involving dark bosons ($X,\Phi$):
\begin{equation} \label{lagxs}
	\mathcal{L}_\text{eff}^{\Phi,X}=\frac{\zeta_\text{eff}}{\Lambda}B_{\mu\nu}X^{\mu\nu}\Phi+\frac{\epsilon_\text{eff}}{\Lambda^2}\bar{\psi}\gamma_\mu\psi \frac{1}{2i}\Phi^\dagger \overleftrightarrow{\mathcal{D}}^\mu\Phi.
\end{equation}
\end{itemize}

\section{Observational limits} \label{sec:obslimits}

We use the following notation for our operators~\footnote{We note the recent study of ref.~\cite{Barman:2020ifq} using OP4, in the context of feebly coupled vector boson DM.}:

\begin{equation} \label{notation}
\begin{aligned}
	\text{OP1} &\equiv B_{\mu\nu}\bar{\Psi}\sigma^{\mu\nu}\Psi,\\
	\text{OP2} &\equiv \bar{\psi}\gamma^\mu\psi\bar{\Psi}\gamma_\mu P_{L,R}\Psi,\\
	\text{OP3} &\equiv B_{\mu\nu}\bar{\Psi}(\gamma^\mu \overleftrightarrow{\mathcal{D}}^\nu-\gamma^\nu \overleftrightarrow{\mathcal{D}}^\mu)P_{L,R}\Psi,\\
	\text{OP4} &\equiv B_{\mu\nu}X^{\mu\nu}\Phi,\\
	\text{OP5} &\equiv \frac{1}{2i}(\bar{\psi}\gamma^\mu\psi)(\Phi^\dagger \overleftrightarrow{\mathcal{D}}_\mu\Phi).
\end{aligned}
\end{equation} 

We also consider the combined contributions from dimension 5 and 6 operators when they contain the same DM candidate; in such cases we adopt the following relationship between the scales $\Lambda$ and operator coefficients $C$:

\begin{equation}
	\Lambda_\text{dim 6}=\Lambda_\text{dim 5}, \hspace{1cm} C_\text{dim 6}= \pm C_\text{dim 5}.
\end{equation}

In most combinations, the relative sign between coefficients is irrelevant, with the exception of the combination between OP1 and OP3, where phenomenology can vary slightly depending on their relative sign.

We are using $\Lambda=2\,m_\text{DM}$ when combining operators of different dimensions \footnote{Although all operators that we consider in this work can, in principle, be generated at tree level by spin-one mediators neutral under both SM and DM gauge groups  \cite{GonzalezMacias:2015rxl}, a caveat is in order. If the dimension 5 operators are generated at loop level, the ratio $m_\text{DM}/\Lambda$ could be a few orders of magnitude smaller. This would depend on the hierarchy between $m_\text{DM}$ and $m_\text{loop}$ (the mass of the inner particle in the loop, not necessarily the mediator or the DM particle), and that is completely model dependent.} \footnote{A comment on the operator coefficients is pertinent: depending on the working assumptions (neutral or charged mediators under SM and DM gauge groups, mediators' spin, etc.) a given operator can be generated at tree level or first appears at one loop (see section 2.1. of \cite{GonzalezMacias:2015rxl}). If the underlying physics is weakly coupled, the coefficient is suppressed by $\sim 1/(16 \pi^2)$, which may require an unnaturally large dimensionless coupling value, that -on the contrary- would be expected if the underlying physics is strongly coupled.
}. We consider that equality a safe limit for the convergence of the effective theory, as discussed in \cite{Busoni:2013lha}. Also in \cite{GAMBIT:2021rlp}, the authors use the same relationship for their calculations to be meaningful in the EFT framework. Depending on the UV completion of the theory, a possible $s$-channel process in the high-energy theory might break the EFT when the corresponding heavy mediator resonates. We have checked that our results change insignificantly moving slightly away from the previous equality ($\Lambda\gtrsim2m_\text{\rm DM}$). Hereafter, we will be expressing  constraints on ratios of effective couplings over $\Lambda^{(2)}$ as bounds on the couplings by using $\Lambda=2\,m_\text{\rm DM}$, for given DM masses.\\
When combining operators, sticking to the case  $\Lambda=2m_{\rm DM}$  maximizes the impact of higher-dimensional operators, through their interference with the leading ones, while keeping the convergence of the EFT. Of course  solutions can be found for $\Lambda>2m_{\rm DM}$. Indeed, as $\Lambda/m_{\rm DM}$ increases, the subleading operators become eventually negligible and the results from the single operators of leading dimension are recovered.

\subsection{Relic density} \label{subsec:RD}
We use micrOMEGAs code \cite{micromegas} to compute the relic abundance of dark matter in our EFT. We use the single operator hypothesis, and we obtain the coefficients in the Lagrangian ---in eqs. (\ref{lagpsi}) and (\ref{lagxs})--- such that they reproduce the observed relic density \cite{Zyla:2020zbs}

\begin{equation} \label{relicdensity}
	\Omega_\text{DM}h^2=0.1200\pm 0.0012\,.
\end{equation}

In the calculations below, we will use the effective couplings that correctly reproduce the relic density, eq. (\ref{relicdensity}). 

\subsection{Direct Detection Experiments} \label{subsec:DD}

\begin{figure}[h!] 
\centering
\includegraphics[width=105mm]{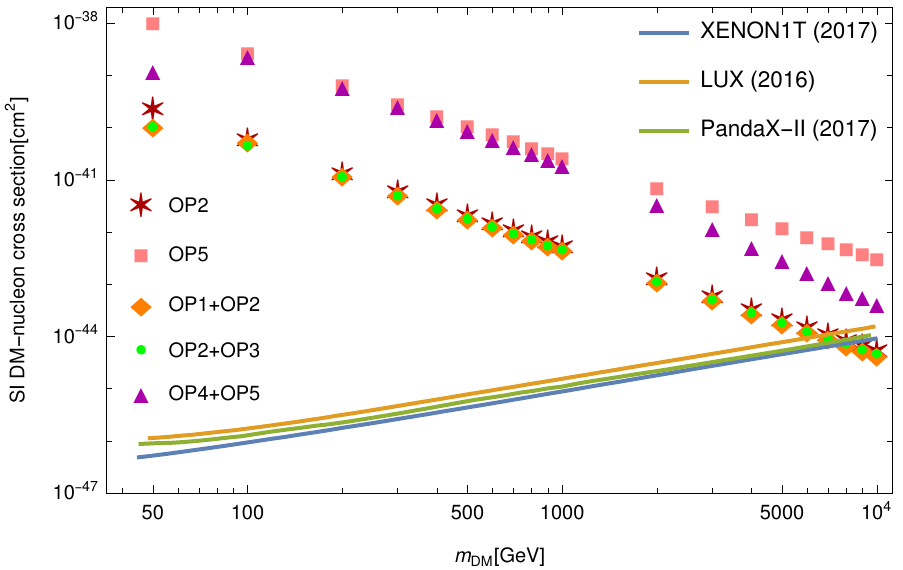}
\caption{WIMP cross sections (normalized to a single nucleon) for spin-independent coupling
versus mass. The notation in this figure is defined in eq. (\ref{notation}). When we combine operators with the same DM candidate, we use $\Lambda = 2 m_\text{DM}$. Operators not shown here have cross sections many orders of magnitude below the current limits.} \label{fig:DD-1}
\end{figure}

For the mass range that we are studying, the most stringent limits on spin-independent scattering cross sections of DM and nucleons come from the LUX-ZEPLIN experiment \cite{LZ:2022ufs}. However, we also include limits from the XENON1T \cite{Aprile:2018dbl} and PandaX-4T \cite{PandaX-4T:2021bab} experiments. Again we use micrOMEGAs \cite{micromegas} to compute the DM-nucleon cross sections within our EFT, in the limit where the relative velocity goes to zero. 
Fig. \ref{fig:DD-1}  shows our results  for several operators in our EFT and compares them with the experimental limits. The notation used in this figure is defined in eq. (\ref{notation}). We can see that OP2, OP5 and the combinations of OP1\&OP2, OP2\&OP3 and OP4\&OP5 are completely ruled out by these experiments. Operators not shown in fig. \ref{fig:DD-1} have DM-nucleon cross sections many orders of magnitude below the current experimental limits from direct detection experiments. Therefore, in the following we will only consider those operators not plotted in fig. \ref{fig:DD-1} ---OP1, OP3, OP4 and the combination of OP1\&OP3---.

\subsection{Dwarf spheroidal satellite galaxies} \label{subsec:DSSG-1}

Using the first year of data from the Dark Energy Survey (DES), eight new dwarf spheroidal satellite galaxies (dSphs) were discovered recently . The dSphs of the Milky Way are some of the most DM dominated objects known. The dSphs are excellent targets for the indirect detection of DM due to their proximity, high DM content, and apparent absence of non-thermal processes. Analyzing Fermi Large Area Telescope data obtained along six years, Ref. \cite{Drlica}  searched for gamma ray emission coincident with the positions of these eight new objects. No significant excess of gamma-ray emission was found.  Then, in Ref. \cite{Drlica} they computed individual and combined limits on the velocity-averaged DM annihilation cross section for these new targets ---assuming that the DES candidates are dSphs with DM halo properties similar to the known dSphs---. 

\begin{figure}[h!] 
\centering
\subfigure[Annihilation into $b\bar{b}$]{\includegraphics[width=80mm]{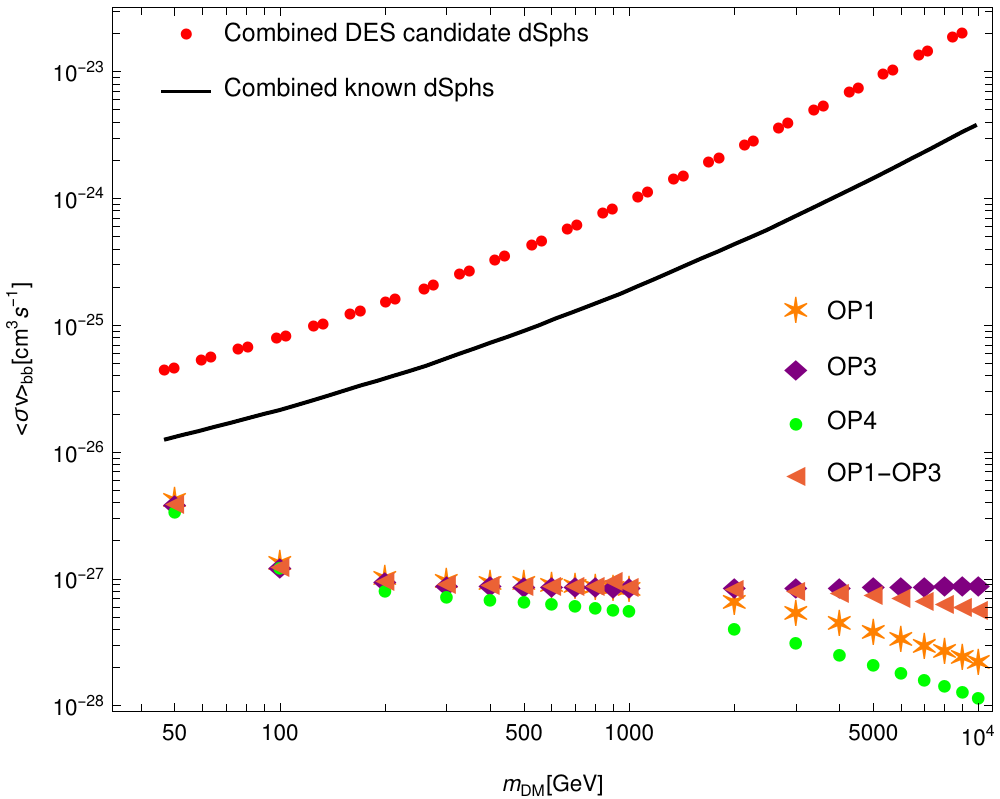}}
\subfigure[Annihilation into $\tau^+\tau^-$]{\includegraphics[width=80mm]{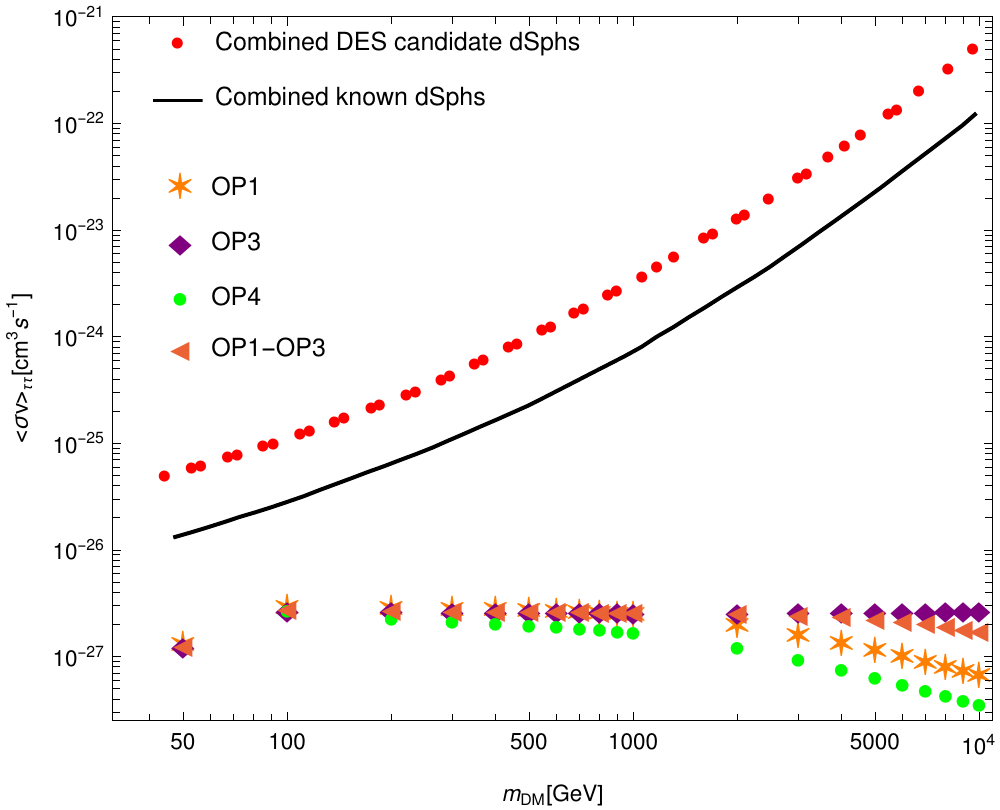}}
\caption{Restrictions from dSphs on the DM annihilation cross sections into (a) $b\bar{b}$, (b) $\tau^+\tau^-$ for the portals generated by several operators, defined in eq. (\ref{notation}). We see in both panels that the entire mass region is allowed by the data.} \label{fig:bb-tt}
\end{figure}

Using micrOMEGAs \cite{micromegas}, we computed the non-relativistic ($m_\text{DM} \ll T$) thermally-averaged DM annihilation cross sections $\left\langle\sigma v\right\rangle$, using our effective operators ---those that are not ruled out by direct detection experiments, see fig. \ref{fig:DD-1}---, and compared the results with the limits mentioned above. The results are presented in figure \ref{fig:bb-tt}, and we can see that these limits do not help us to constrain our mass region. Note that the combination of operators OP1 and OP3 has a relative sign between its coefficients, because the one with the same sign gives velocity-averaged cross sections even below those shown in the figures.

\subsection{Limits from AMS-02 positron measurements} \label{subsec:AMS-1}

\begin{figure}[h!] 
\centering
\subfigure[Annihilation into $e^+e^-$]{\includegraphics[width=80mm]{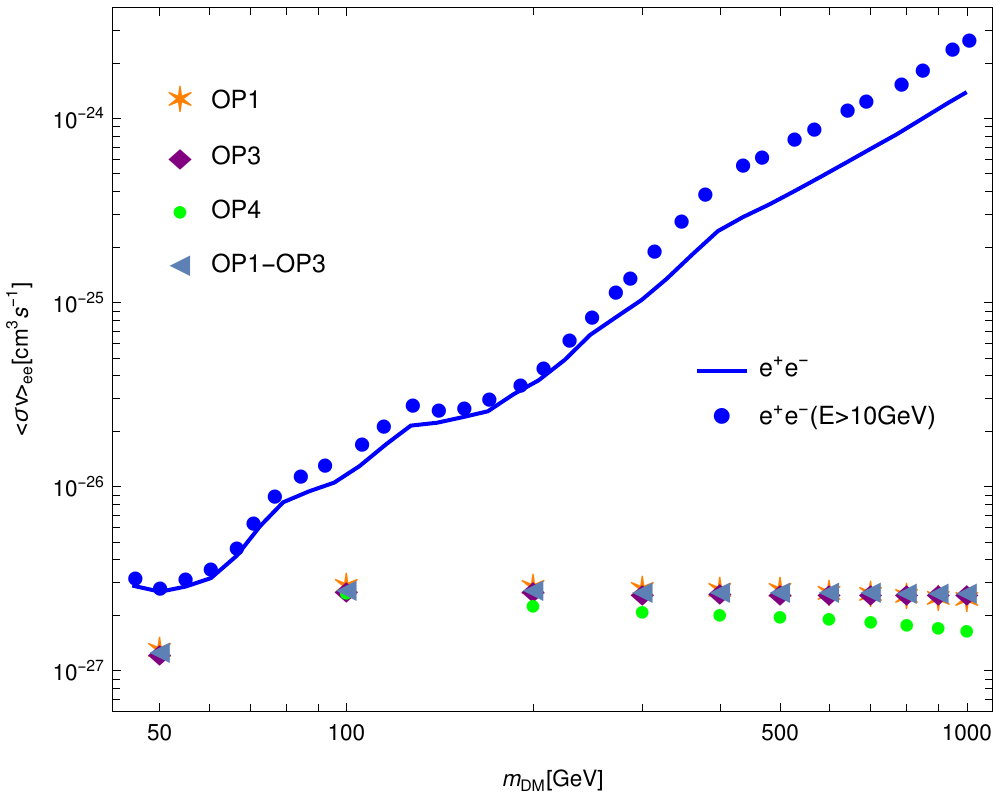}}
\subfigure[Annihilation into $\mu^+\mu^-$]{\includegraphics[width=80mm]{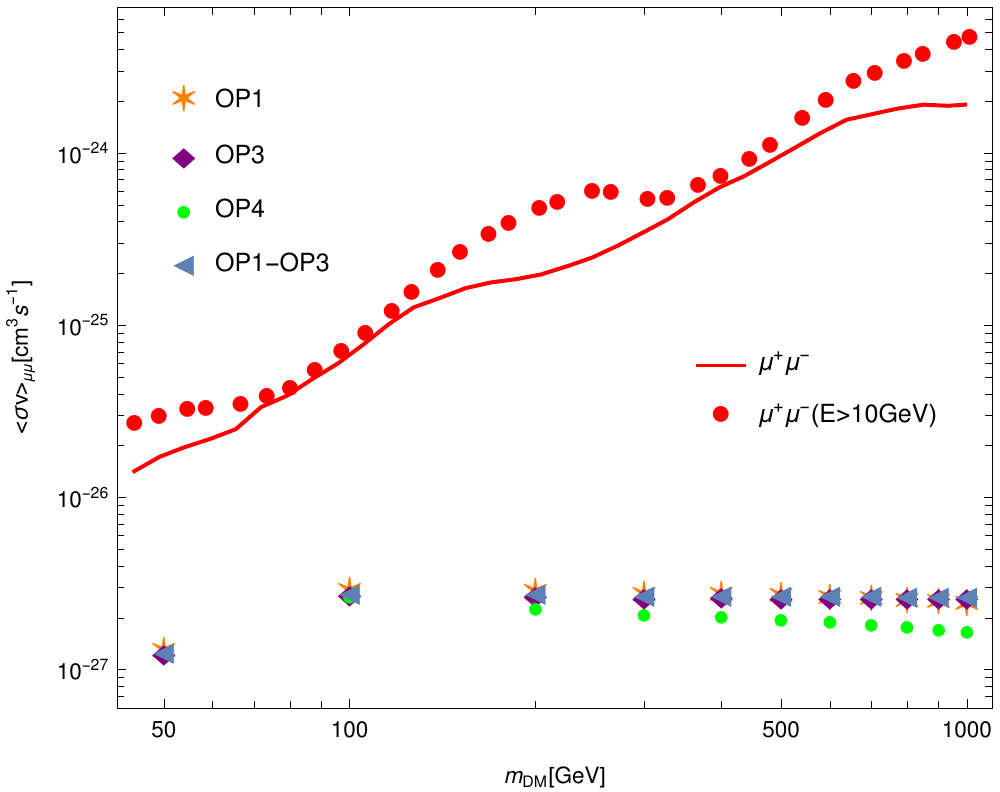}}
\caption{Restrictions from AMS-02 data on the DM annihilation cross sections into (a) $e^+e^-$ and (b) $\mu^+\mu^-$ for the portals generated by several operators, defined in eq. (\ref{notation}). We see that the entire mass region is allowed by the data. The limits shown as solid lines were derived from sampling over various energy windows, while the dashed lines are from considering those windows including only data with energies above 10 GeV \cite{Ibarra:2013zia}.} \label{fig:ee-mm}
\end{figure}

The AMS-02 Collaboration has presented high-quality measurements of positron fluxes as well as the positron fraction. In Ref. \cite{Ibarra:2013zia} the authors used measurements of the positron flux to derive limits on the dark matter annihilation cross section and lifetime for various final states, and extracted strong limits on DM properties. They worked under the well-motivated assumption that a background positron flux exists from spallations of cosmic rays with the interstellar medium and from astrophysical sources. We again computed the DM annihilation cross sections, now into $e^+e^-$ and $\mu^+\mu^-$, using micrOMEGAs \cite{micromegas} and compare them with the bounds derived in Ref. \cite{Ibarra:2013zia}. They also derived limits for the $\tau^+\tau^-$ and $b\bar{b}$ final states, but these are weaker than those from dSphs data. In figure \ref{fig:ee-mm} we see that our results are below the experimental limits and we cannot rule out any mass region. Note that in this figure we again show the combinations of OP1 and OP3 with a relative sign between their coefficients, while their combination with the same sign gives even smaller values for the velocity-averaged cross sections.

\begin{figure}[h!] 
\centering
\includegraphics[width=100mm]{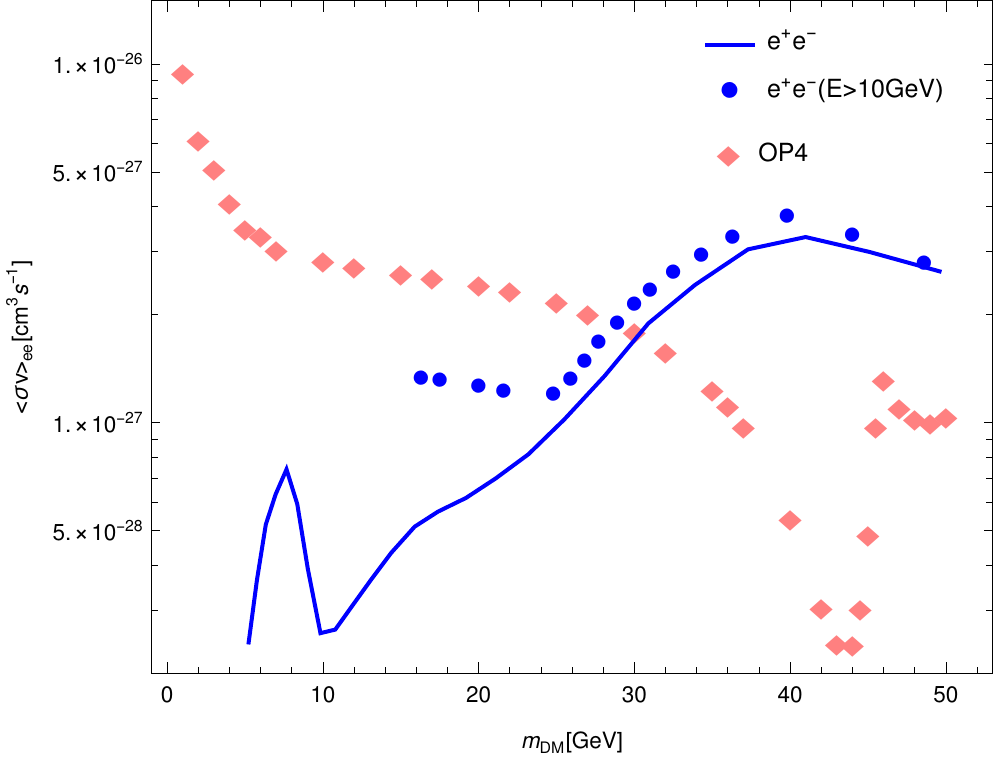}
\caption{Restrictions from AMS-02 data on the DM annihilation cross sections into $e^+e^-$ for the portal generated by OP4, defined in eq. (\ref{notation}). This plot tests the mass region $m_\psi<m_Z/2$, and we see that masses larger than $\sim 30$ GeV are allowed.} \label{fig:e-OP4}
\end{figure}

We refine our calculation of the DM annihilation cross sections done previously in ref. \cite{Fortuna:2020wwx} \footnote{Before, we used the first two terms of a series expansion of $\langle\sigma v\rangle$ as a function of $x=m/T$, where $m$ stands for the DM mass and $T$ is the temperature. In this work we used micrOMEGAs to compute $\langle\sigma v\rangle$ more accurately (the updated values are shown in fig. \ref{fig:e-OP4}). This change explains the small difference in the low mass region of OP4, between the results summarized in tables \ref{tab:res} and \ref{tab:res-1}.} and the region of masses allowed was slightly modified. This change is only noteworthy in the case of the OP4, because the collider constraints exclude masses in the region $m_\psi < m_Z/2$ for OP1, OP3 and the combinations of OP1 \& OP3, as we will see below. The data constraining DM annihilation into the final state $e^+e^-$ is the most stringent, therefore is the one we present here, in fig. \ref{fig:e-OP4}. We see that masses smaller than $\sim 30$ GeV are ruled out, while masses in the range $[30,50]$ GeV are allowed.


\subsection{Is the EFT perturbative?}
We want our EFT to be in the perturbative regime, which imposes an upper limit in the dimensionless effective couplings of eqs. (\ref{lagpsi}) and (\ref{lagxs}). We bind it using that the corresponding $\alpha=g^2/(4\pi)$, where $g^2$ stands for any coupling in eqs. (\ref{lagpsi}) and (\ref{lagxs}), should be at most $\sim 1/2$ to keep perturbativity. As before, we took the effective couplings that correctly reproduce the relic abundance. We found that, for the OP4 with $m_X\neq m_\Phi$: if the smaller mass is $<1$ TeV, relationships as $m_X=3\,m_\Phi$, $m_\Phi=3\,m_X$ are allowed. While if the smaller mass is \mbox{$1$ TeV $<m<
 3.2$ TeV}, the other particle can only be twice heavier. The quantities that we obtained for the rest of the operators satisfy this criterion of perturbativity.

\section{Collider constraints} \label{sec:CollCons}

The effective operators we are working with allow for the pair production of WIMPs ($\chi$) in the proton–proton collisions at the LHC. If one of the incoming partons radiates a jet through initial state radiation (ISR), one can observe the process $p p\to\chi\chi j$ as a single jet associated with missing transverse energy ($\slashed{E}_T$). In this study, we include the ATLAS \cite{ATLAS:2021kxv} monojet analysis based on $139\,\text{fb}^{-1}$ of data from Run II. ATLAS has performed a number of further searches for other types of ISR, leading for example to mono-photon signatures, but these are known to give weaker bounds on DM EFTs than monojet searches \cite{Zhou:2013fla, Brennan:2016xjh, Bauer:2017fsw}.

Starting from UFO files generated using LanHEP v4.0.0 \cite{Semenov:2014rea}, we have then generated the process $p p\to\chi\chi j$ with MadGraph\_aMC@NLO v3.4.0 \cite{Alwall:2014hca} for the ATLAS analysis, interfaced to Pythia v8.3 \cite{Bierlich:2022pfr} for parton showering and hadronization. The detector response is simulated using the ATLAS detector configuration \cite{Araz:2020lnp} in FastJet v3.3.3 \cite{Cacciari:2011ma}. We apply the following kinematic cuts from Ref.~\cite{ATLAS:2021kxv}: $E_T^{\rm miss}>200$ GeV, a leading jet with $p_T>150$ GeV and $|\eta|<2.4$,  and up to three additional jets with $p_T>30$ GeV and $|\eta|<2.8$.

\begin{figure}[htb]
    \centering
    \includegraphics[width=11cm]{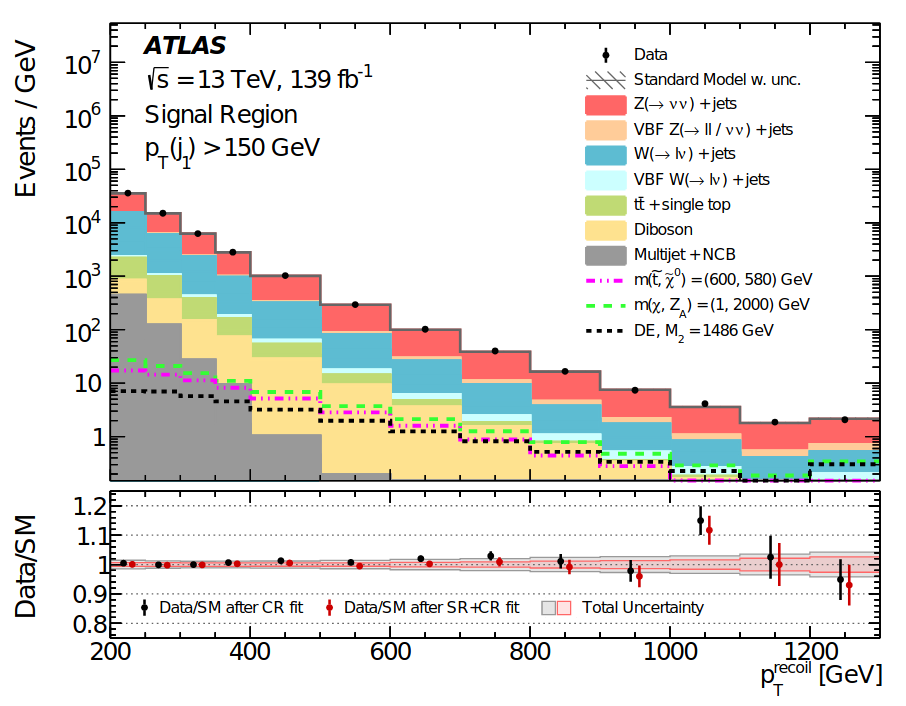}
    \caption{Measured distributions of $p_T^\text{recoil}$ for $p_T^\text{recoil}>200$ GeV selection \cite{ATLAS:2021kxv} compared with the SM predictions in the signal region.}
    \label{fig:ATLAS}
\end{figure}

We validated our analysis by reproducing the green dash-dotted line in figure \ref{fig:ATLAS}, using a simplified DM model where Dirac fermion WIMPs ($\chi$) are pair-produced from quarks via $s$-channel exchange of a spin-1 mediator particle ($Z_A$) with axial-vector couplings~\cite{ATLAS:2021kxv}.

In this analysis we only include the operators (and combinations of them) that still had mass regions with suitable solutions ---OP1, OP3, OP4 and the combinations of OP1\&OP2 and OP1\&OP3---, allowed even after all the constraints imposed by non-collider experiments that we have considered. The results reported by ATLAS were obtained using proton-proton collision data at a center-of-mass energy of $\sqrt{s}=13$ TeV. Events were required to have at least one jet with transverse momentum above $200$ GeV and no reconstructed leptons or photons. Due to the $\sqrt{s}=13$ TeV center-of-mass energy, the maximum mass we considered in our simulations was $6.4$ TeV. We use the data points in fig. \ref{fig:ATLAS} of the measured distributions of $p_T^\text{recoil}$. 

Renormalization group effects can  produce a sizable running of the Wilson coefficients between the low-energy scales probed in direct detection experiments and the high energies of the LHC (see \cite{Bishara:2016hek, Bishara:2017nnn, Bishara:2017pfq, Bishara:2018vix} and references therein), which depend on $\mathcal{O}_{\rm SM}$, see eq.~(\ref{eq:factorized_op}). For our operatos in eqs.~(\ref{lagpsi}) and (\ref{lagxs}), QCD effects are negligible. We disregarded QED mixing affecting $A_{\mathrm{eff}}^{L/R}$ (its corresponding operator, OP2, is excluded in the entire region studied).

\section{Discussion and Conclusions} \label{sec:Concl-1}

We recall that operators OP2, OP5 and the combinations OP1+OP2, OP2+OP3 and OP4+OP5 \footnote{The combinations of OP1\&OP2 and OP2\&OP3 are ruled out mainly due to the contribution of OP2 to the spin-independent DM-nucleon cross sections, which does not exclude OP1 or OP3 alone. Similarly, the combination of OP4\&OP5 is excluded mostly due to the contribution of OP5 to the SI DM-nucleon cross sections, which does not exclude OP4 alone.} were already excluded in the range $[50\text{ GeV},6.4\text{ TeV}]$ by direct detection experiments data. We show below the results obtained by comparing the data from ATLAS \cite{ATLAS:2021kxv} (see fig. \ref{fig:ATLAS}) with the simulated results for each operator. When we combined operators, for every benchmark point evaluated in the simulations, the relation $\Lambda=2\, m_{\rm DM}$ was used.

\begin{table}[h!]
  \begin{center}
    \begin{tabular}{|l|c|c|c|} 
	  \hline
	  \rowcolor{lightcyan}     
      \textbf{Operator} & \textbf{Dim.} & \textbf{DM candidate} & \textbf{{\small Allowed DM mass (GeV)}} \\
      \hline
      1.- $B_{\mu \nu} \bar{\Psi} \sigma^{\mu \nu} \Psi$ & 5 & $\Psi$ fermion & $\approx 0.0025-2,\approx 33-44.5$ \\
      \rowcolor{cosmiclatte}      
      2.- $\left(\bar{\psi} \gamma_\mu \psi \right) \left(\bar{\Psi} \gamma^\mu P_{L,R} \Psi \right)$ & 6 & $\Psi$ fermion & none \\
      3.- {\small $B_{\mu \nu} \bar{\Psi} (\gamma^\mu \protect\olra{\mathcal{D}}^\nu - \gamma^\nu \protect\olra{\mathcal{D}}^\mu) P_{L, R} \Psi$} & 6 & $\Psi$ fermion & $\approx 33-44.5$ \\
   	  \rowcolor{cosmiclatte}      
      4.- $B_{\mu \nu} X^{\mu \nu} \Phi$ & 5 & {\small vector $X$, scalar $\Phi$} & $\approx 0.11-2, \approx 36-44.5$ \\
      5.- $\left(\bar{\psi} \gamma_\mu \psi \right)\, \frac{1}{2i} \Phi^\dagger \overleftrightarrow{\mathcal{D}}^\mu \Phi$ & 6 & scalar $\Phi$ & none \\
   	  \rowcolor{cosmiclatte}      
      $\qquad\qquad 1\pm2$ & 5+6 & $\Psi$ fermion & $\approx 0.0025-2$ \\
      $\qquad\qquad1\pm3$ & 5+6 & $\Psi$ fermion & $\approx 0.0025-2, \approx 33-44.5$ \\
   	  \rowcolor{cosmiclatte}      
      $\qquad\qquad2\pm3$ & 6 & $\Psi$ fermion & none \\     
      \hline
    \end{tabular}
  \end{center}
  \caption{Summary of results obtained in ref. \cite{Fortuna:2020wwx}: considering the $Z$ invisible decay width, relic density, direct detection experiments and indirect detection results from dSphs and positron flux measurements. It is very important to note that we are considering masses of the dark particles below the mass of the $Z$ boson ($M_Z/2\sim 45. 5$ GeV, as they appear in charge conjugated pairs). } \label{tab:res}      
\end{table}

We also complemented our previous results from Ref. \cite{Fortuna:2020wwx} \footnote{In Ref. \cite{Fortuna:2020wwx} we select benchmark values for $\Lambda$ ($230$ GeV $<\Lambda< 1$ TeV), but when we combined operators, its value was irrelevant.}, shown in table \ref{tab:res}, so we tested the solutions found according to the experimental data analyzed there, in the region $m_\text{DM}<m_Z/2$.
. We show below the comparison of the simulated events, for masses previously allowed, with the ATLAS data. 
\begin{figure}[htb]
    \centering
    \includegraphics[width=8cm]{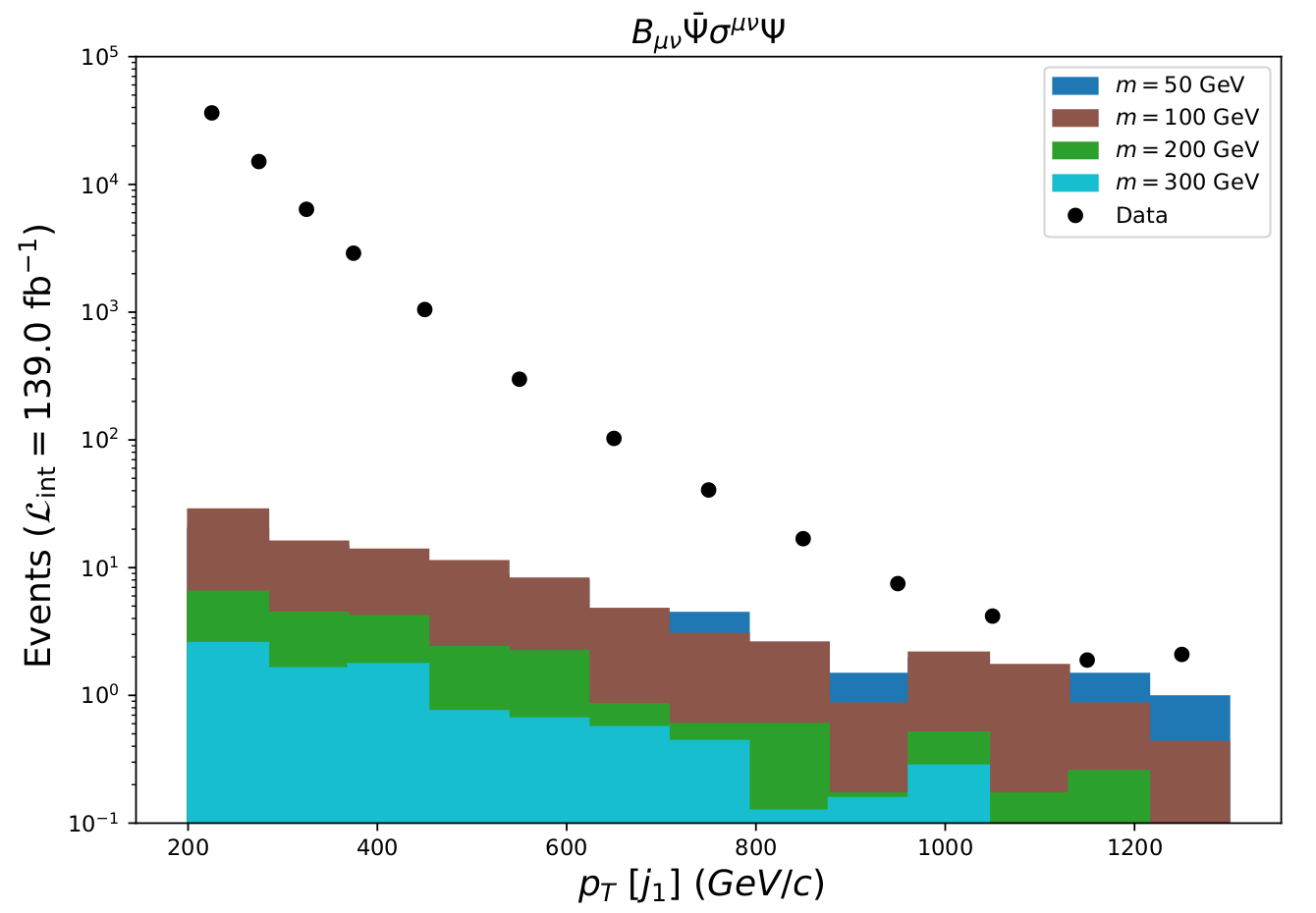}
    \caption{$p_T$ distributions simulated using OP1 of eq. (\ref{notation}), vs ATLAS data (fig. \ref{fig:ATLAS}). We use benchmark points for $50$ GeV, $100$ GeV, $200$ GeV and $300$ GeV. We see that all these masses are allowed.}
    \label{fig:OP1-tev}
\end{figure}

\begin{figure}[htb]
\centering
\subfigure[]{\includegraphics[width=8cm]{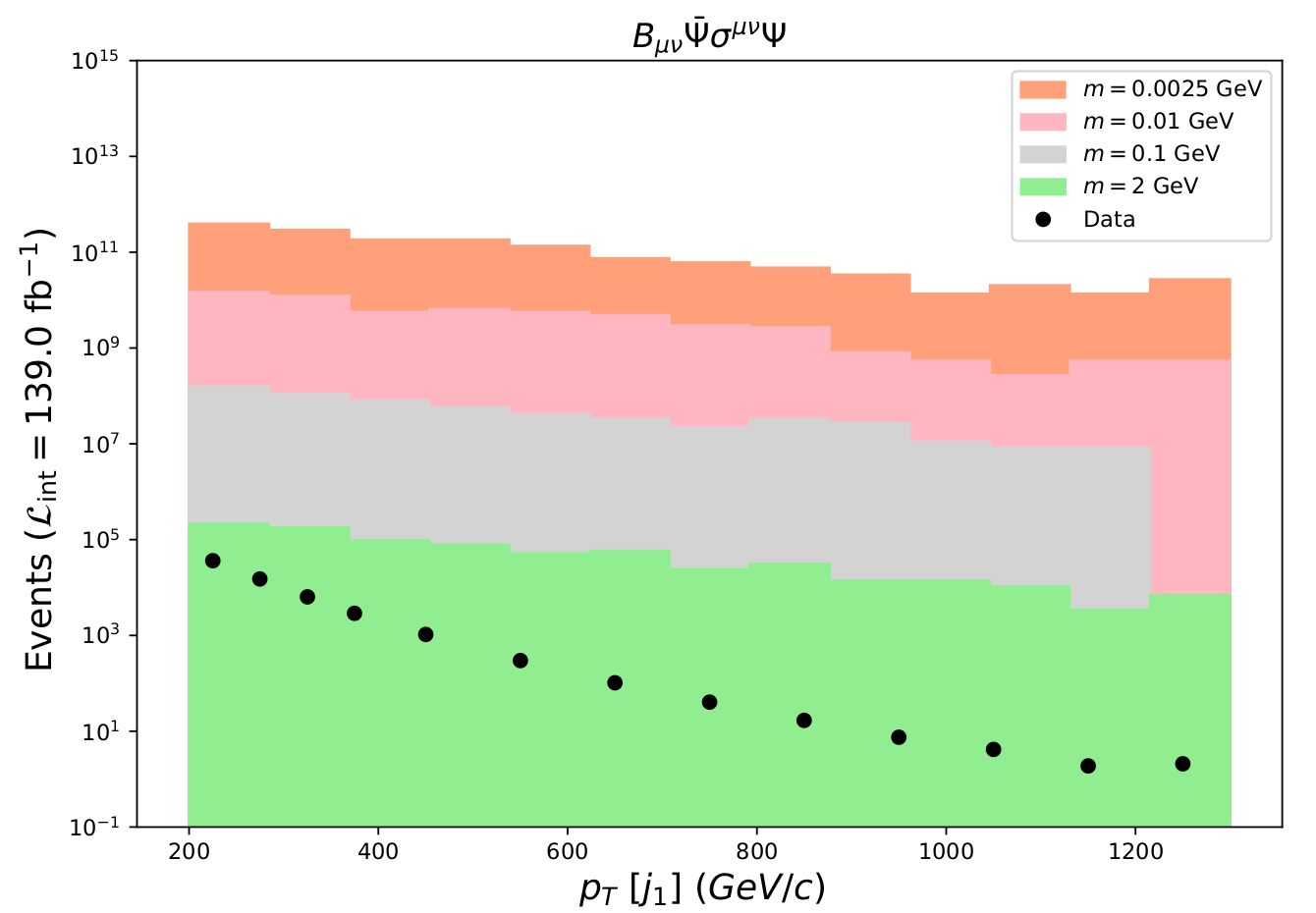}} 
\subfigure[]{\includegraphics[width=8cm]{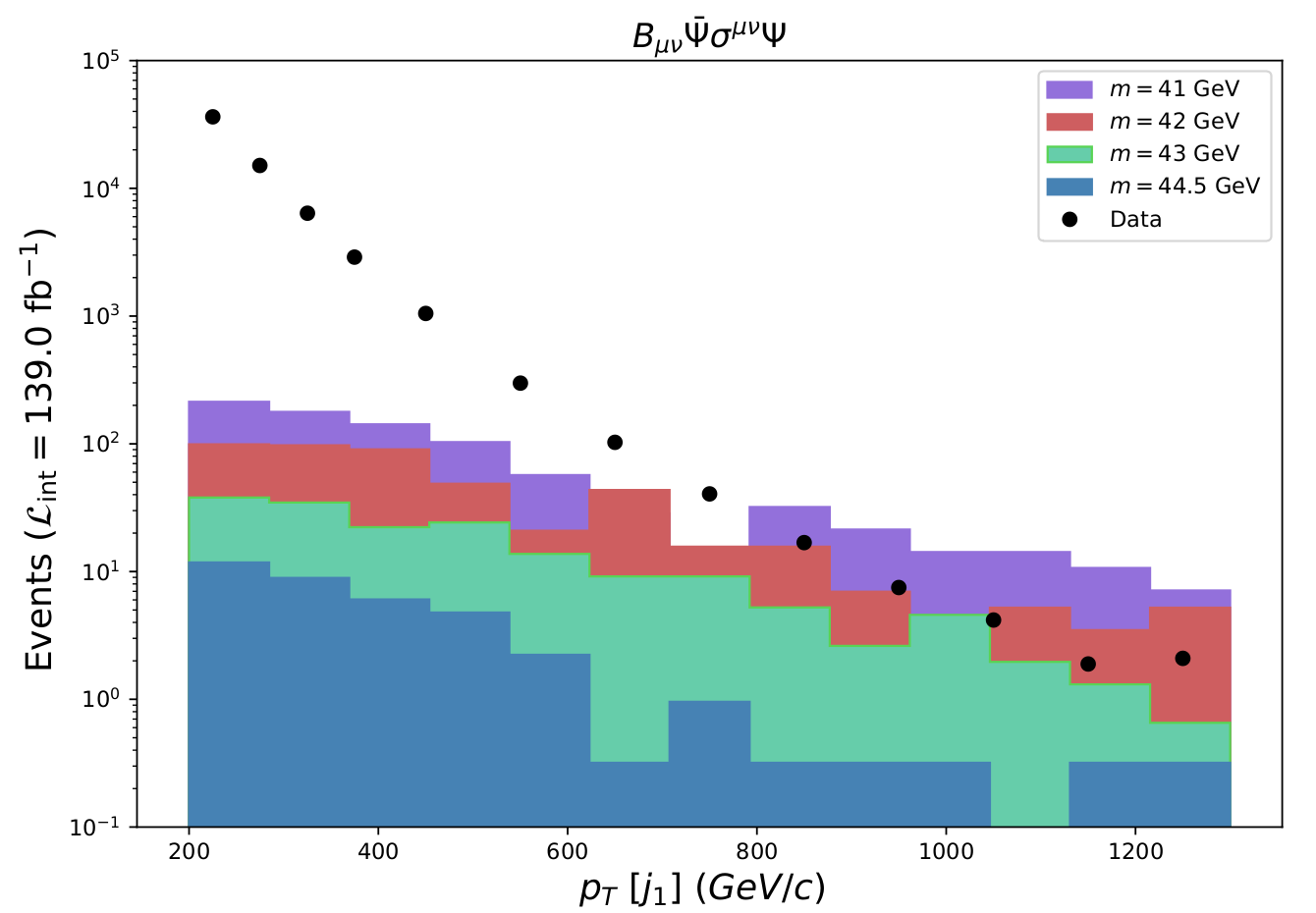}}
\caption{$p_T$ distributions simulated using OP1 of eq. (\ref{notation}), vs ATLAS data (fig. \ref{fig:ATLAS}). We use benchmark points for (a) $0.0025$ GeV, $0.01$ GeV, $0.1$ GeV and $2$ GeV, and (b) $41$ GeV, $42$ GeV, $43$ GeV and $44.5$ GeV. We see that masses smaller than 43 GeV are excluded by the data.} \label{fig:OP1-mz}
\end{figure}

\begin{figure}[htb]
\centering
\subfigure[]{\includegraphics[width=8cm]{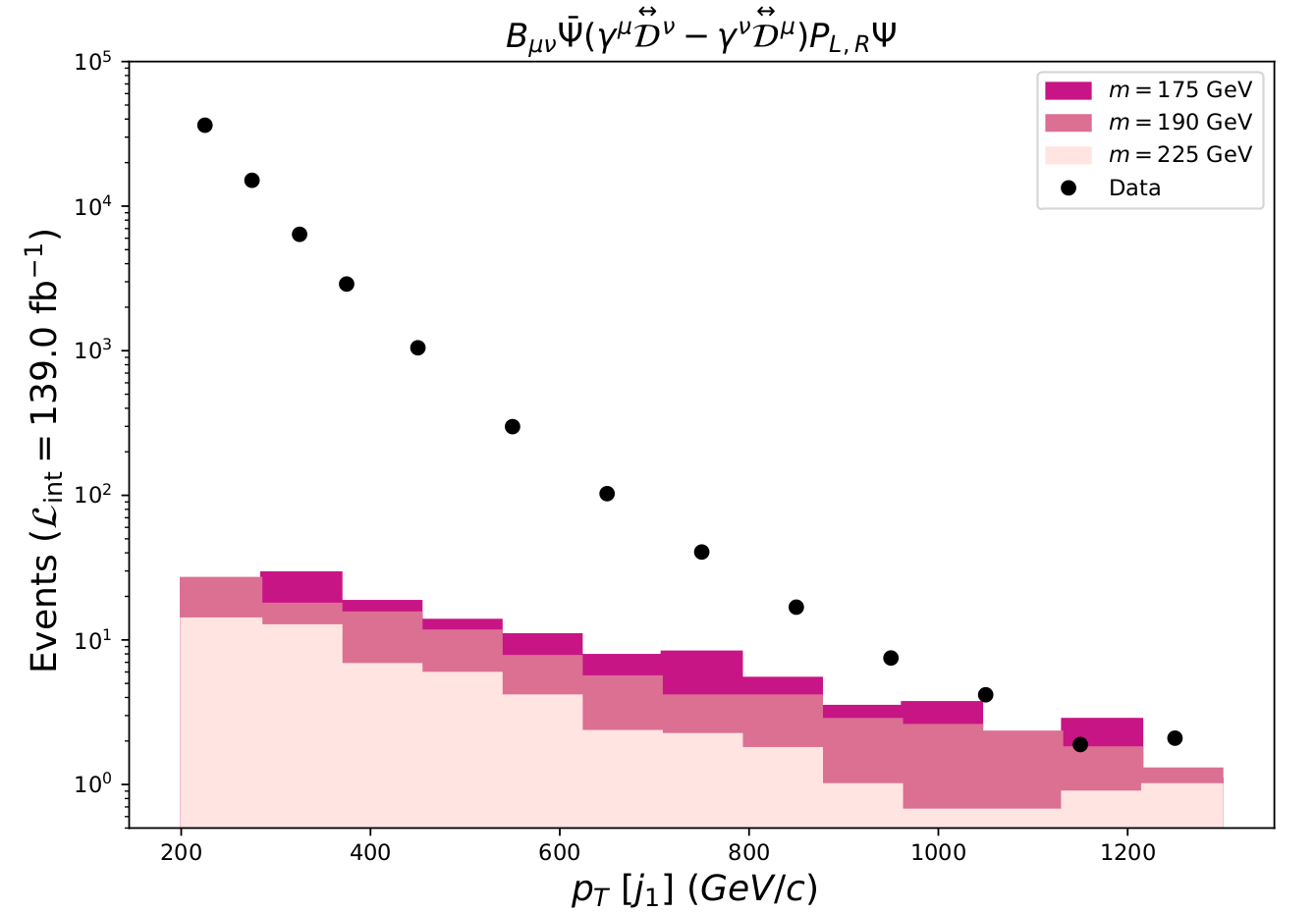}}
\subfigure[]{\includegraphics[width=8cm]{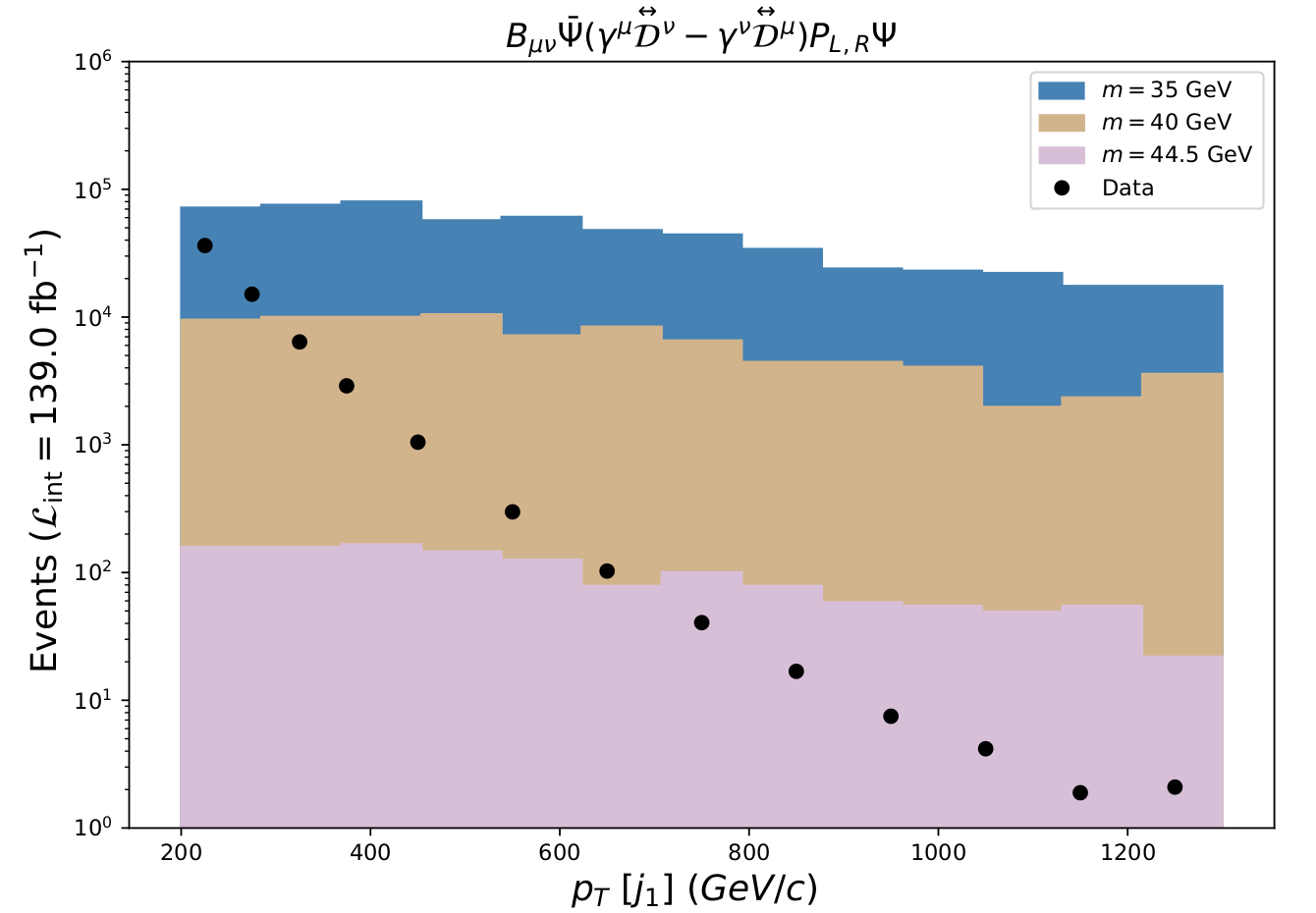}}
\caption{$p_T$ distributions simulated using OP3 of eq. (\ref{notation}), vs ATLAS data (fig. \ref{fig:ATLAS}). We use benchmark points for (a) $175$ GeV, $190$ GeV and $225$ GeV and (b) $35$ GeV, $40$ GeV and $44.5$ GeV. The plot in (a) shows that masses above $190$ GeV are allowed, while in (b) we see that all the region is excluded.} \label{fig:OP3}
\end{figure}

\begin{figure}[htb]
\centering
\subfigure[]{\includegraphics[width=7.8cm]{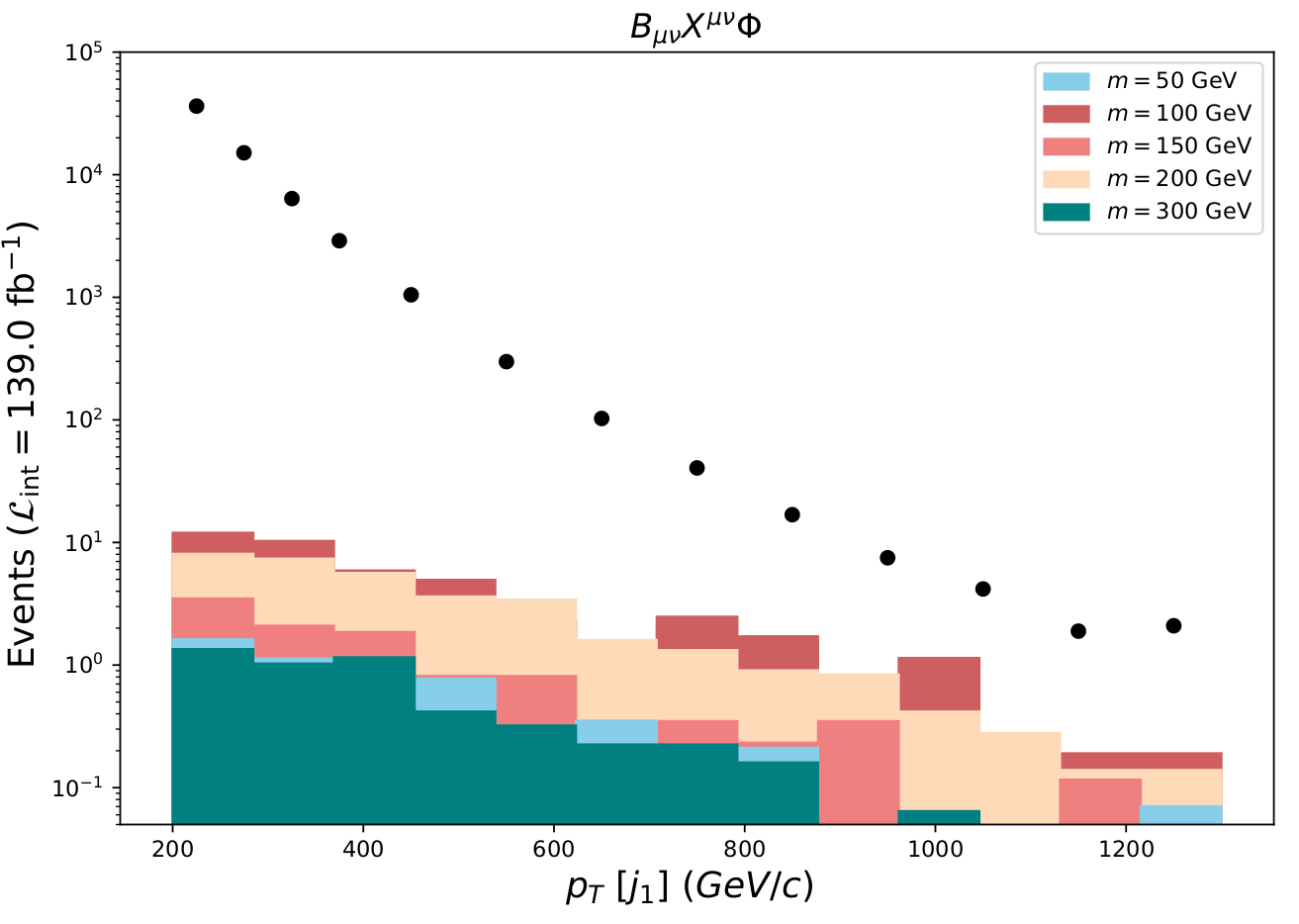}}
\subfigure[]{\includegraphics[width=8cm]{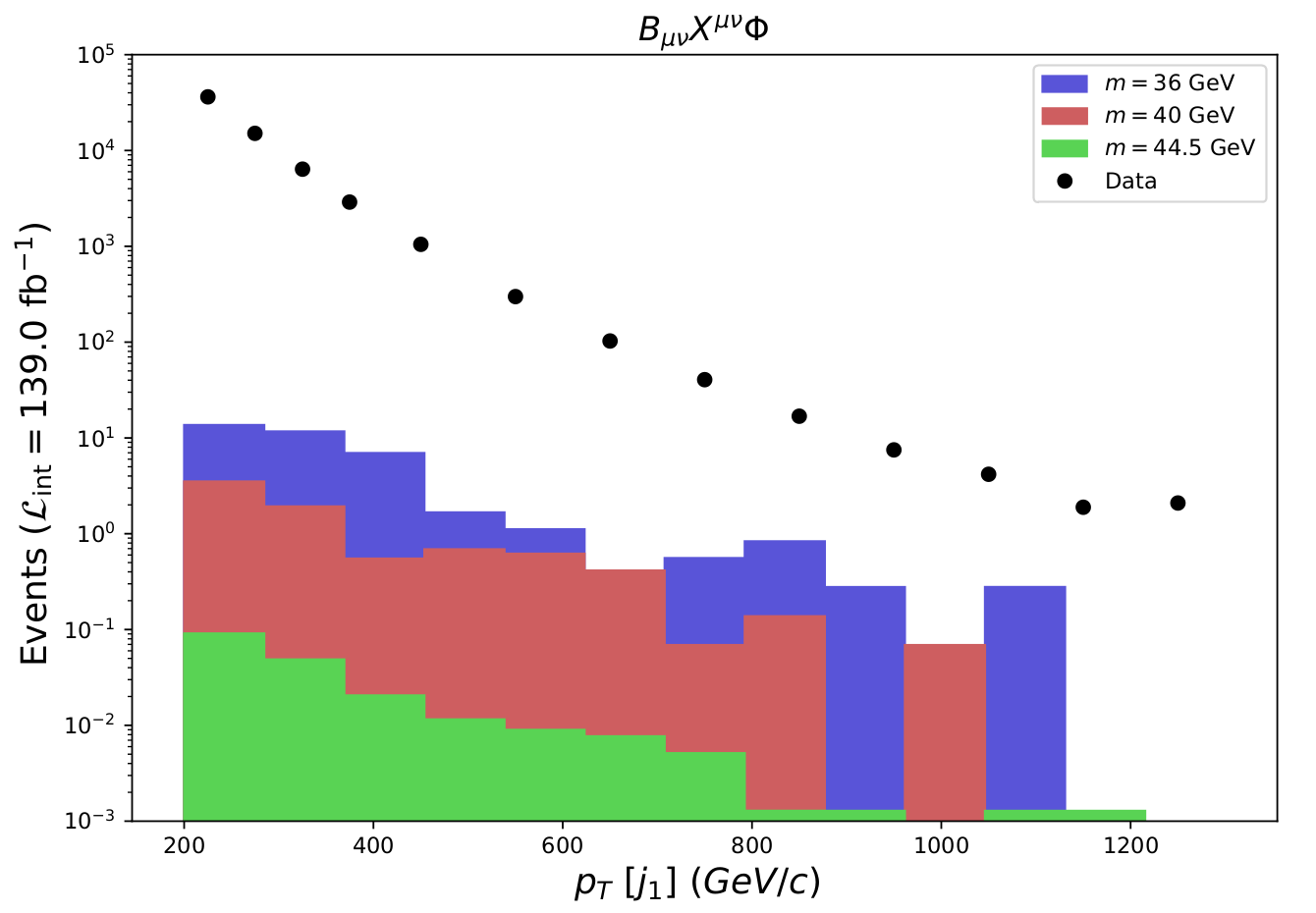}}
\caption{$p_T$ distributions simulated using OP4 of eq. (\ref{notation}), vs ATLAS data (fig. \ref{fig:ATLAS}). We use benchmark points for (a) $50$ GeV, $100$ GeV, $150$ GeV, $200$ GeV and $300$ GeV and (b) $36$ GeV, $40$ GeV and $44.5$ GeV. We see that all these masses are allowed.} \label{fig:OP4}
\end{figure}

\begin{figure}[htb]
    \centering
    \includegraphics[width=8cm]{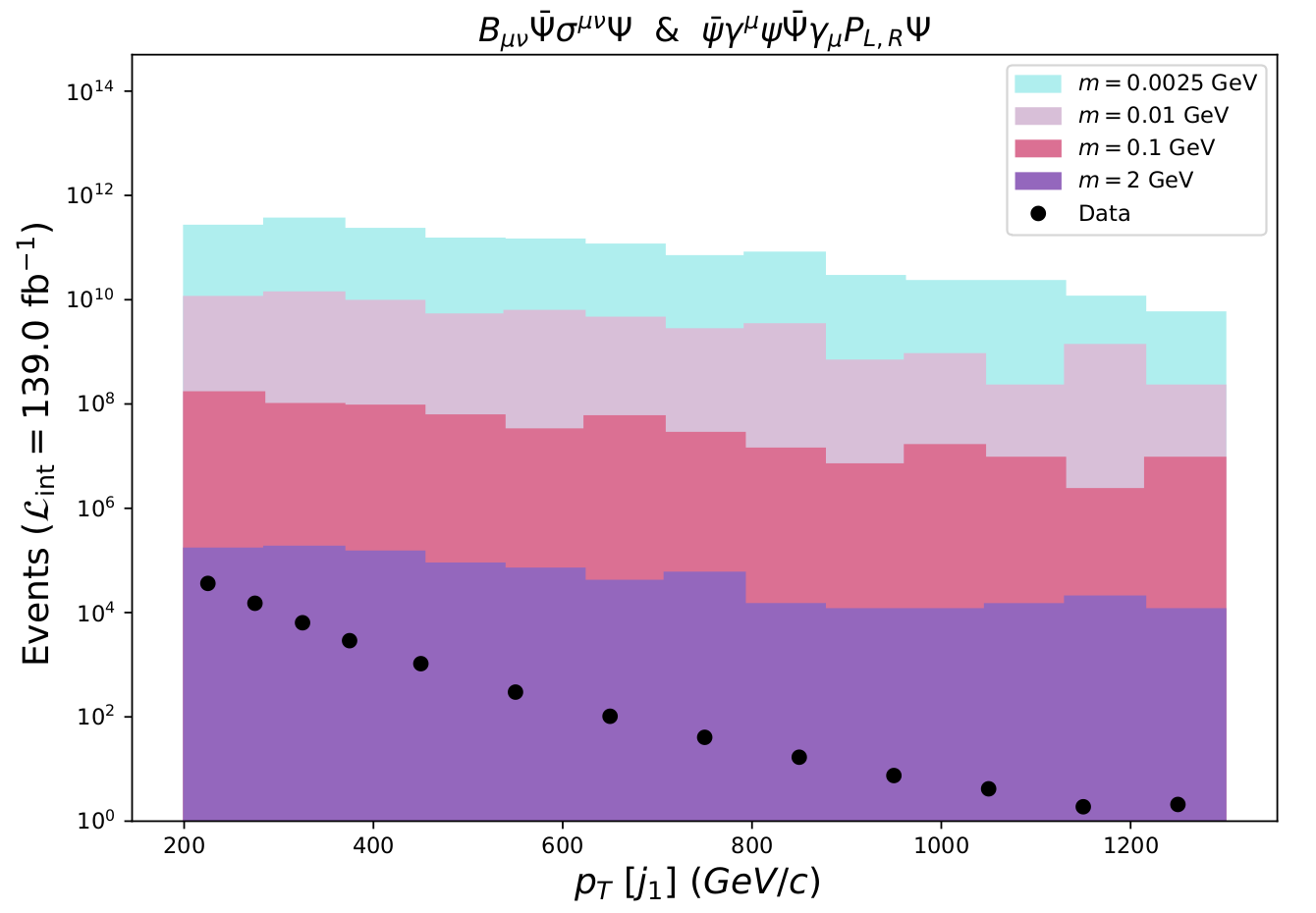}
    \caption{$p_T$ distributions simulated using OP1\&OP2 of eq. (\ref{notation}), vs ATLAS data (fig. \ref{fig:ATLAS}). We use benchmark points for $0.0025$ GeV, $0.01$ GeV, $0.1$ GeV and $2$ GeV. We see that all these masses are ruled out.}
    \label{fig:OP1y2}
\end{figure}

\begin{figure}[htb]
\centering
\subfigure[OP1+OP3]{\includegraphics[width=8cm]{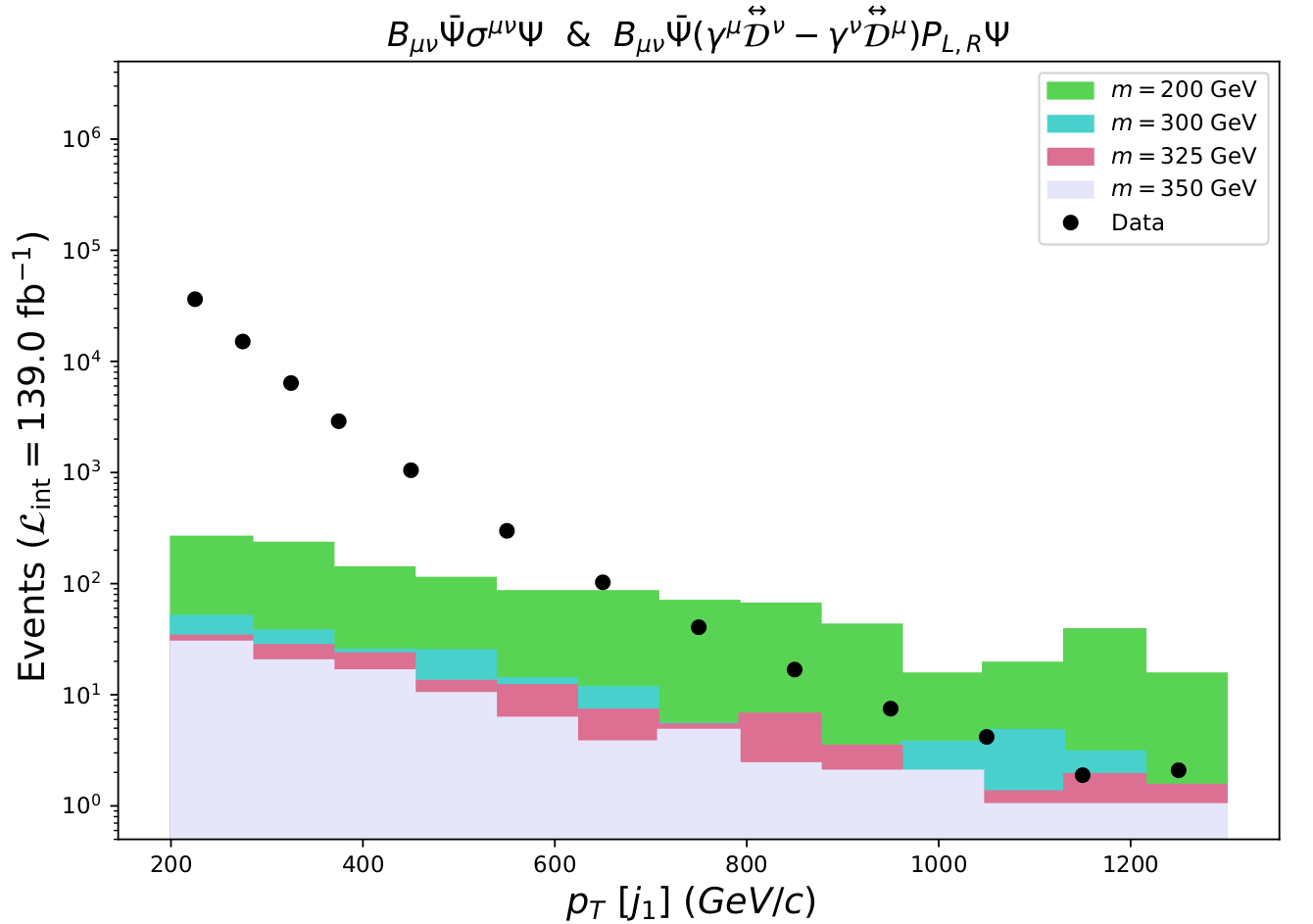}}
\subfigure[OP1-OP3]{\includegraphics[width=8cm]{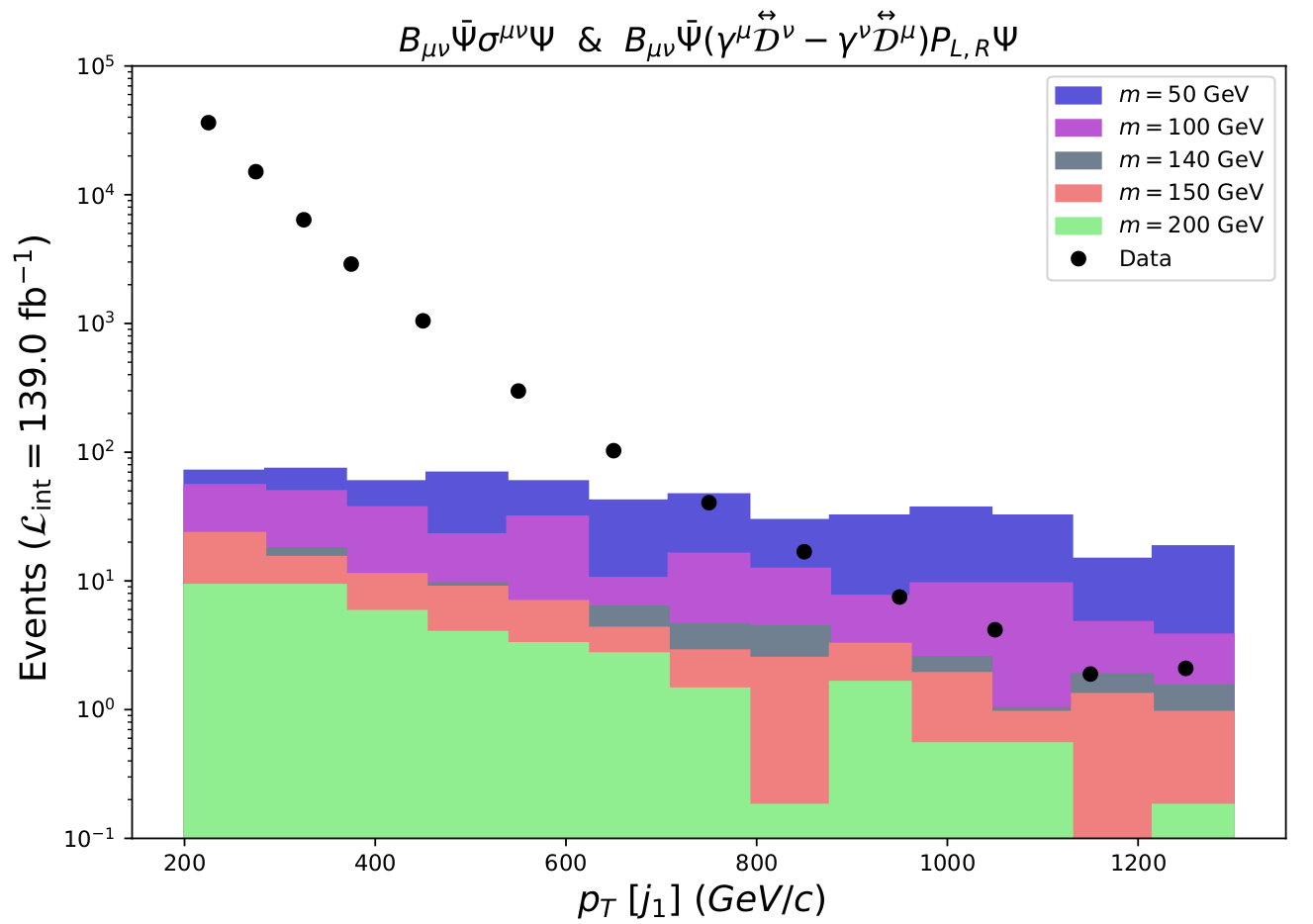}}
\caption{$p_T$ distributions simulated using (a) OP1+OP3 and (b) OP1-OP3 of eq. (\ref{notation}), vs ATLAS data (fig. \ref{fig:ATLAS}). We use benchmark points for (a) $200$ GeV, $300$ GeV, $325$ GeV and $350$ GeV and (b) $50$ GeV, $100$ GeV, $140$ GeV, $150$ GeV and $200$ GeV. The masses allowed are (a) above $325$ GeV and (b) above $140$ GeV.} \label{fig:OP1y3}
\end{figure}

\begin{figure}[htb]
\centering
\subfigure[]{\includegraphics[width=8cm]{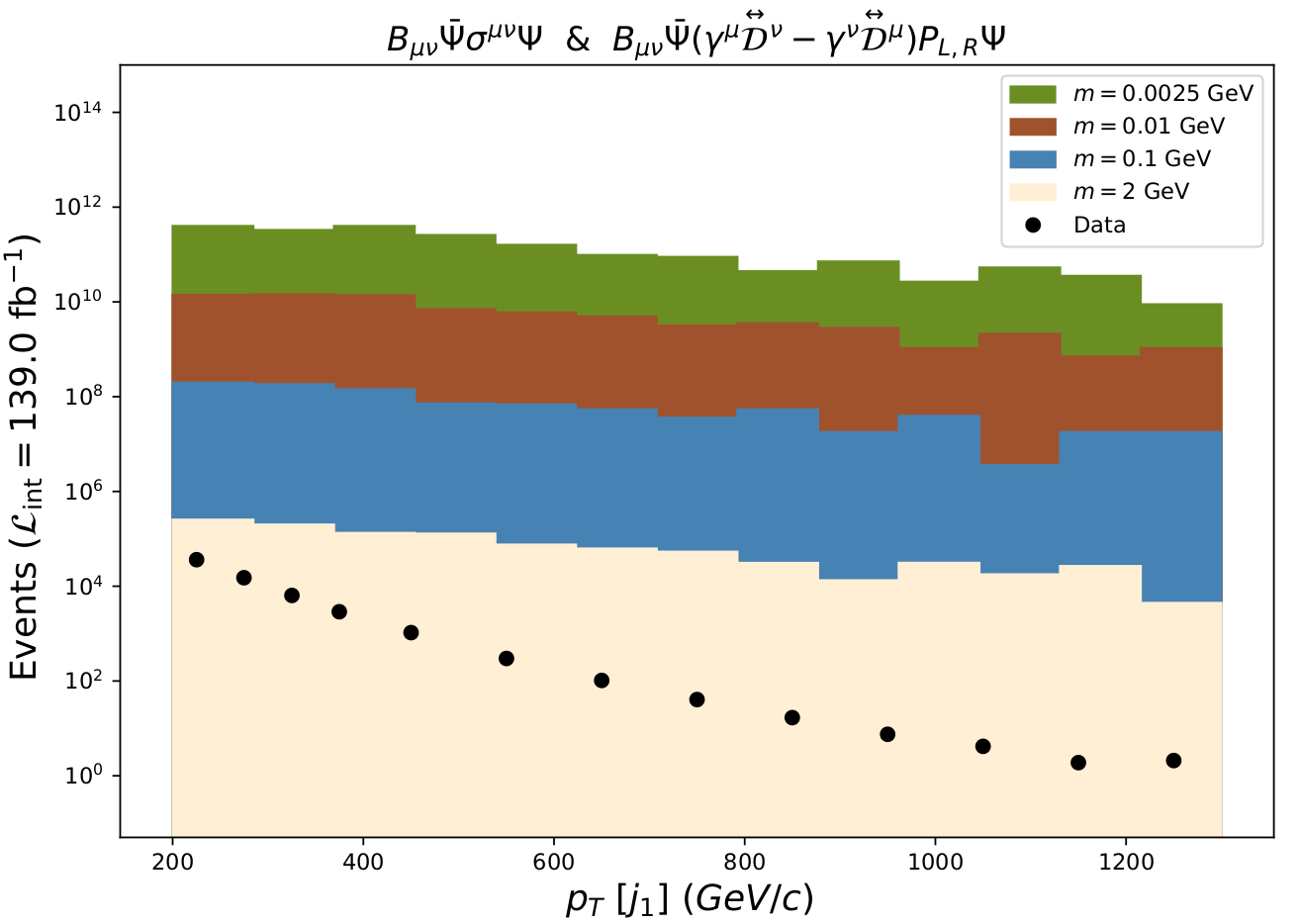}}
\subfigure[]{\includegraphics[width=8cm]{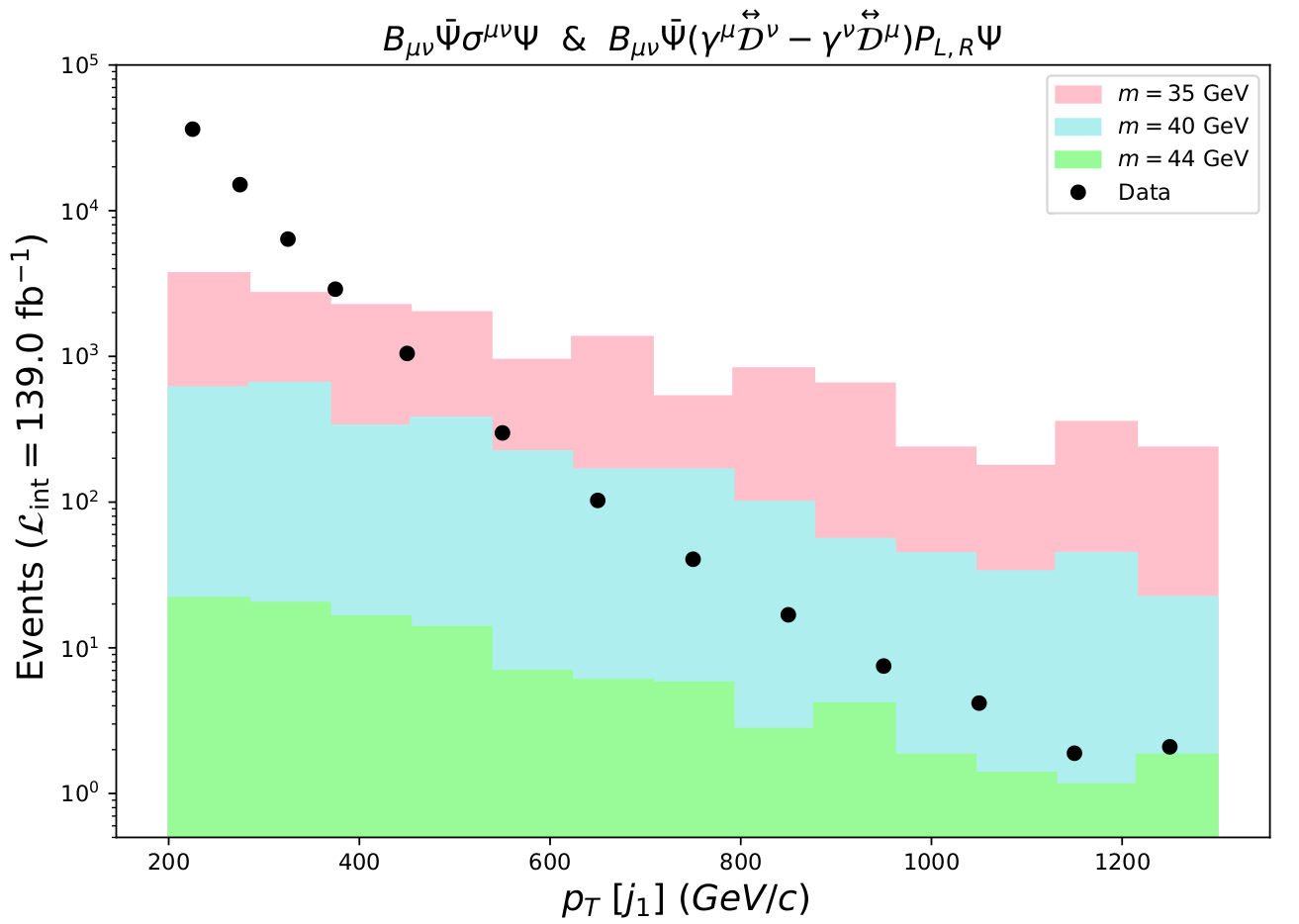}}
\caption{$p_T$ distributions simulated using OP1 \& OP3 of eq. (\ref{notation}), vs ATLAS data (fig. \ref{fig:ATLAS}). We use benchmark points for (a) $0.0025$ GeV, $0.01$ GeV, $0.1$ GeV and $2$ GeV and (b) $35$ GeV, $40$ GeV and $44$ GeV. In (a) all the masses are ruled out by the data, while in (b) masses larger than $44$ GeV are allowed.} \label{fig:OP1y3-mz}
\end{figure}

\begin{itemize}
    \item OP1. In fig. \ref{fig:OP1-tev} we evaluated $m_\psi= 50$ GeV, $100$ GeV, $200$ GeV and $300$ GeV and we observe that all these masses are allowed. In fig. \ref{fig:OP1-mz} we use the benchmark points (a): $0.0025\; {\rm GeV}, 0.01\; {\rm GeV}, 0.1\; {\rm GeV\; and\;} 2\; {\rm GeV}$, and (b) $41\; {\rm GeV}, 42\; {\rm GeV}\;, 43\; {\rm GeV\; and\;} 44.5\; {\rm GeV}.$ We see that masses smaller than $43$ GeV are ruled out for this operator. 
    \item OP3. We evaluated the masses: $175\; {\rm GeV}, 190\; {\rm GeV\; and\;} 225\; {\rm GeV}$ in fig.~\ref{fig:OP3}(a) and $35\; {\rm GeV}, 40\; {\rm GeV}$ and $44.5\; {\rm GeV}$ in fig.~\ref{fig:OP3}(b). We see in fig.~\ref{fig:OP3}(a) that masses larger than $190$ GeV are allowed. For this operator the region $m_\psi<M_Z/2$ is now entirely excluded.
    \item OP4. In fig. \ref{fig:OP4} we use the benchmark points: (a) $50$ GeV, $100$ GeV, $150$ GeV, $200$ GeV and $300$ GeV and (b) $100$ GeV, $36$ GeV, $40$ GeV and $44.5$ GeV. We see in both figures that all values are allowed by the data.
    \item OP1\&OP2. We use benchmark points for $0.0025$ GeV, $0.01$ GeV, $0.1$ GeV and $2$ GeV in fig. \ref{fig:OP1y2}. We see that all these masses are excluded by the data.
    \item OP1\&OP3. We evaluated the masses: (a) $200$ GeV, $300$ GeV, $325$ GeV and $350$ GeV, and (b) $50$ GeV, $100$ GeV, $140$ GeV, $150$ GeV and $200$ GeV. In fig. \ref{fig:OP1y3}(a) we use the same sign for the effective  couplings and in fig. \ref{fig:OP1y3}(b) we use a relative sign between the operators. The masses allowed are (a) larger than $325$ GeV and (b) larger than $140$ GeV. Finally, for DM masses below $M_Z/2$, we tested the benchmark points: fig. \ref{fig:OP1y3-mz}(a) $0.0025$ GeV, $0.01$ GeV, $0.1$ GeV and $2$ GeV and fig. \ref{fig:OP1y3-mz}(b) $35$ GeV, $40$ GeV, $44$ GeV. We see that in both figures  the whole mass range is ruled out by the data ($m_{\rm DM}\in [44\; {\rm GeV},M_Z/2]$ was already excluded by analysis of positron measurements, see table \ref{tab:res}).
\end{itemize}    
    
We present a summary of our results in table \ref{tab:res-1}.

\begin{table}[h!]
  \begin{center}
    \begin{tabular}{|l|c|c|c|} 
	  \hline
	  \rowcolor{lightcyan}     
      \textbf{Operator} & \textbf{Dim.} & \textbf{DM candidate} & \textbf{{\small Allowed DM mass}} \\
      \hline
      1.- $B_{\mu \nu} \bar{\Psi} \sigma^{\mu \nu} \Psi$ & 5 & $\Psi$ fermion & $\gtrsim 43$ GeV $^*$ \\
      \rowcolor{cosmiclatte}      
      2.- $\left(\bar{\psi} \gamma_\mu \psi \right) \left(\bar{\Psi} \gamma^\mu P_{L,R} \Psi \right)$ & 6 & $\Psi$ fermion & none \\
      3.- {\small $B_{\mu \nu} \bar{\Psi} (\gamma^\mu \protect\olra{\mathcal{D}}^\nu - \gamma^\nu \protect\olra{\mathcal{D}}^\mu) P_{L, R} \Psi$} & 6 & $\Psi$ fermion & $\gtrsim 190$ GeV\\
   	  \rowcolor{cosmiclatte}      
      4.- $B_{\mu \nu} X^{\mu \nu} \Phi$ & 5 & {\small vector $X$, scalar $\Phi$} & $\gtrsim 30$ GeV $^*$\\
      5.- $\left(\bar{\psi} \gamma_\mu \psi \right)\, \frac{1}{2i} \Phi^\dagger \overleftrightarrow{\mathcal{D}}^\mu \Phi$ & 6 & scalar $\Phi$ & none \\
   	  \rowcolor{cosmiclatte}      
      $\qquad\qquad 1\pm2$ & 5+6 & $\Psi$ fermion & none \\
      $\qquad\qquad1+3$ & 5+6 & $\Psi$ fermion & $\gtrsim 325$ GeV \\
   	  \rowcolor{cosmiclatte}      
      $\qquad\qquad1-3$ & 5+6 & $\Psi$ fermion & $\gtrsim 140$ GeV \\
      $\qquad\qquad2\pm3$ & 6 & $\Psi$ fermion & none \\     
      \hline
    \end{tabular}
  \end{center}
  \caption{Summary of results obtained in this work, which supersede those in our previous paper  \cite{Fortuna:2020wwx}. In addition to the experimental constraints used therein, now we also considered the limits from ATLAS in ref. \cite{ATLAS:2021kxv}. $^*$We note that the region $(M_Z\pm\Gamma_Z)/2$ is excluded, see Table~\ref{tab:res}.} \label{tab:res-1}      
\end{table}

The constraining power of ATLAS results forbids mostly light DM particles with masses below $M_Z/2$. For OP1 and OP4, we still have solutions below $M_Z/2$, while for OP3 and the combination of OP1\&OP3 we need larger masses to satisfy the ATLAS constraints. Future LHC analyses will set even tighter constraints on DM, particularly within our EFT and, specifically, for the subset of operators (those with spin-one mediators) considered in this work. 

\section*{Acknowledgements}We are grateful to José Wudka for valuable discussions and suggestions and to Marco A. Arroyo-Ureña for his helpful guide and advice with the software used in this work. F.~F.~ acknowledges financial support from CONACyT graduate grants program
No. 728500. P. R is indebted to C\'atedras Marcos Moshinsky (Fundaci\'on Marcos Moshinsky) and CONACyT 
(‘Paradigmas y Controversias de la Ciencia 2022’, Project No. 319395).

\bibliographystyle{unsrt}
\bibliography{biblio}

\end{document}